\documentclass[ms,11pt]{egs}
  \usepackage{times}                   % WITH TIMES ROMAN FONT
\usepackage{graphicx}
\printfigures              % PRINTS OUT FIGURES AT END OF MANUSCRIPT AND
\begin{document}

\title{Scaling in the space climatology of the auroral indices: Is SOC the only possible description ?}  
\author{N. W. Watkins }
\affil{
 British Antarctic Survey, High 
Cross, Madingley Rd., Cambridge CB3 0ET, UK
}
\date{\today}
\journal{\NPG}
\msnumber{12345}

\maketitle
\begin{abstract}
 The study of the robust features of the magnetosphere is motivated both by
new  ``whole system" approaches, and by the idea of ``space climate"
 as opposed to ``space weather". We enumerate these features for the
 $AE$ index, and discuss whether self-organised criticality (SOC) is the most natural explanation
 of the ``stylised facts" so far known for $AE$.
 We  identify and discuss some open questions, answers to which will clarify the extent to
 which $AE$'s properties provide evidence for SOC.
  We then suggest an  SOC-like reconnection-based
 scenario drawing on the result of
 \cite{craig} as an explanation of the very recent 
 demonstration by \cite{uritskyjgr} of power laws in
 several properties of spatiotemporal features
seen in auroral images.  
\end{abstract}

%\narrowtext
%\twocolumn
\section{Introduction: The ``stylised facts" of the $AE$ indices}
In the last few years,  the advent of magnetospheric
 models (e.g. \cite{chang1999,consolini1997,chapman1998}) based on the idea 
 of self-organised criticality (SOC) (e.g. \cite{jensenbook,sornette}) or more general critical
phenomena has stimulated new efforts towards the identification of those features of the magnetosphere
which are robust and repeatable in statistical studies taken over arbitrarily
long periods of time. This new knowledge therefore provides
potentially important information for studies of magnetospheric 
``space climate" (e.g. \cite{boteler}), as distinct from  the many  case studies of individual 
magnetospheric events, in particular magnetic storms and substorms, a field 
increasingly referred to as ``space weather". 
SOC models typically have power law probability density functions (PDFs) 
and ``1/f" power spectra (PSDs) for energy release from the system, 
signals of a multiscale process with no preferred scale in space or time.
 
 In mathematical economics, the robust and repeatable mathematical properties
 of an economic index or other variable are sometimes referred to as ``stylised
 facts" (e.g. \cite{mikosch}). In this paper we seek to establish some of the stylised facts of the magnetosphere,
 and to sketch out a possible type of simple model for some of its outputs.
  We have  chosen the 
specific example of  the $AE$ indices ($AE, AU$, $AL$ and $AO$).
 Our purpose is twofold:

Firstly we wish to collect some still-relevant results on the space climatology of the indices which were obtained before the 
rise of the SOC paradigm (see also the earlier review of \cite{klimas}). New
 findings in this area  tend now to be exclusively interpreted within a  framework such
as SOC or other nonequilibrium critical phenomena such as noise assisted topological
phase transitions. It seems to us, however, that such behaviour will be most clearly
demonstrable if an alternative framework is developed in parallel that
seeks to encompass the data but does not require avalanching or self-regulation 
(see also \cite{freeman2000b,watkins2001b,watkins2001c}). If and when such a description fails,
 the manner in which it does so would make much more convincing the need for an SOC (or more generally 
 a self-regulating  description). Insofar as it works, however, such a description is a useful mathematical tool
 for quantifying the risk of extremes in the indices, and possibly also in
 mathematically  analogous solar  wind quantities.
 
  Secondly, we wish to reiterate our reasons for thinking that  the  scale-free behaviour of the auroral indices need
not necessarily have the same origin as the scale free behaviour seen in newer observations,
most notably those based on ultraviolet images. This is important because
the modelling of the indices is of continuing space weather/climate interest, but should neither be restricted by the  SOC paradigm
or be a restriction on the construction of magnetospheric SOC models unless 
it is actually necessary.

It is worth stressing that such arguments are by no means unique to the present author, or
the topic of magnetospheric physics. \cite{fs} have recently given a clear statement of the need for the construction of 
such null models, commenting on a claim of SOC in queues in the UK National Health Service.
They illustrate that in this case identical behaviour to that observed is also seen in 
(non self-regulating) Brownian motion. More generally, a very clear discussion of the extent to
which self-regulating models are needed to explain the ubiquitous scale-free
phenomena in natural and man-made systems  has been given by \cite{sornette}
(see also his recent summary \citep{sornette2002}). In the plasma context \cite{krommes} has given an 
explicit counter-example to the idea that power law tails in time-domain  correlation functions
are necessarily evidence of SOC. The present paper is an attempt to make some initial
developments of this approach in magnetospheric physics, extending some of the ideas first expressed
in \cite{watkins2001c} and \cite{chapman2000}.

 The plan of the paper is as follows: Partly because they have been extensively used as  proxies for {\bf energy} dissipation, 
and thus can be treated as dynamical variables in chaos or SOC-inspired
studies, the attributes of the $AE$ indices have been widely studied. In Section 1 we  present a critical summary of what has been
learned about the indices. We discuss whether SOC is necessary to account for the
 stylised facts of the indices.  In particular we make the 
 suggestion that AE may be satisfactorily
 explained as a solar-wind driven component 
  modelled by a  multiplicative processes, such as fractional lognormal 
 motion, if coupled with a second component, describing the intrinsic ``unloading"
 behaviour.  In Section 2 we discuss several questions that
we believe need to be answered more fully in order to clarify the  most
appropriate model for $AE$. In particular
we note that systems with lognormal amplitude distributions, when also
exhibiting time domain persistence, will tend to
give  power law-like PDFs for the burst
measures previously applied to AE, without the presence of self-organisation,
or even truly scale-free amplitude behaviour. We suggest that this
kind of ``persistent lognormality" is a suitable null model against which
SOC should be tested in the auroral indices, and may well have more widespread
relevance. In Section 3 we discuss newer auroral imager-derived measurements which promise to identify
SOC behaviour more unambiguously, and we  draw attention to the existence of a 
magnetic reconnection-based physical  model which may explain some aspects of these most recent results.
 Section 4
gives our conclusions.  Because of the deliberately narrow focus of this
paper we have not dealt with all of the available literature on SOC in the
magnetosphere. In consequence the reader is encouraged to consult the review papers
by \cite{chapman2000} and \cite{consolini2000},  and also the extended discussions of the evidence for SOC and/or
SOC-like multiscale behaviour in the magnetosphere given by \cite{chang1999},
 \cite{klimas1999}, \cite{lui2002}  and \cite{sitnov2001} 

\subsection{Time series and PDF}
$AE$  estimates maximal ionospheric current densities via the upper ($AU$) and lower ($AL$) envelopes of magnetic 
perturbations and  is usually derived  from 12 magnetometers underneath 
the  mean position of the  auroral electrojets. 
The total envelope $AE=AU-AL$. In figure 1a we show an example of three days of the 
$AE$ time series in early 1975, as studied by \cite{consoliniprl}.
As noted by \cite{consoliniprl}, the first difference of $AE$ (figure 1b) shows 
apparent ``burstiness". However a  fractional lognormal motion (flm) - a fractional 
Brownian motion (e.g. \cite{malamud}) with a lognormal rather than Gaussian
distribution of amplitudes  -  would also show such burstiness. More significantly
the first difference of $\log_e AE$ is also bursty, whereas an flm would show
a normal distribution in that quantity (see figure 6). This suggests that a single fractional
lognormal motion would fail as a model of $AE$.

\begin{figure}
  \includegraphics[width=15cm,height=10cm]{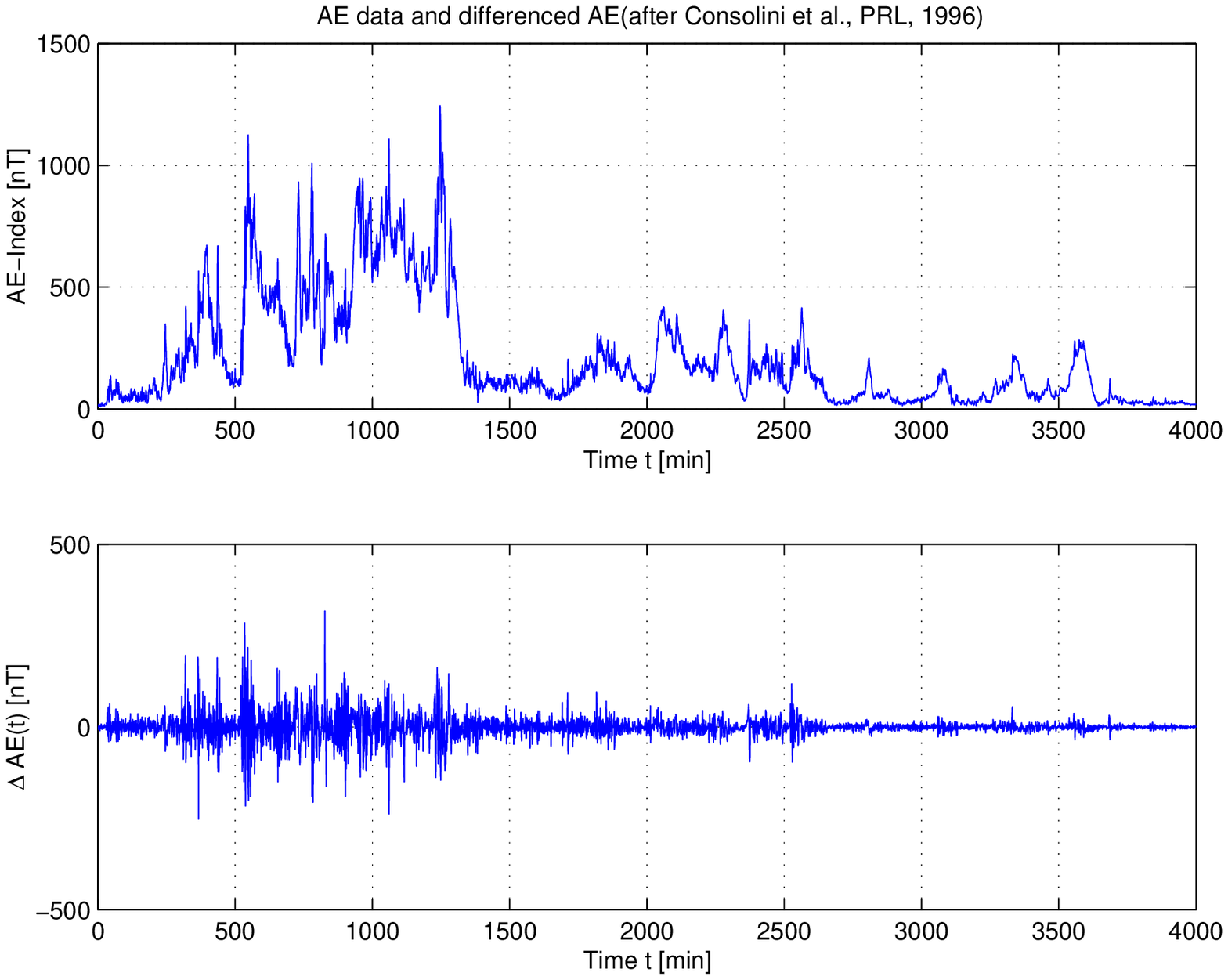}
  \includegraphics[width=15cm,height=10cm]{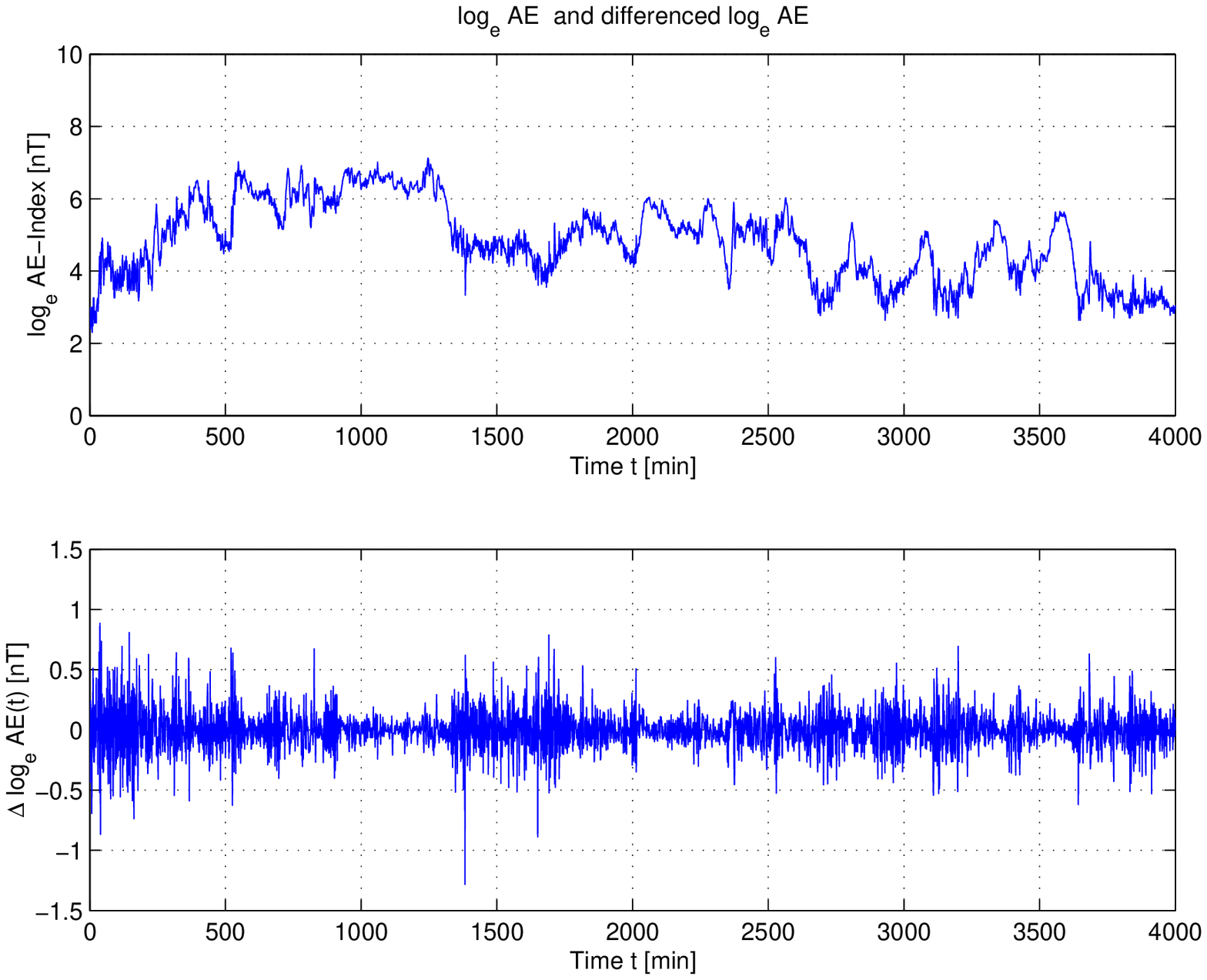}
\caption{The top panel (Figure 1a) shows $AE$ for 4000 minutes in 1975, 
as plotted by \cite{consoliniprl}. The second panel (Figure 1b) shows the first differences
of the data, while the third and fourth panels (Figures 1c and 1d) show the natural logarithm of
$AE$ and its first difference. The impulsiveness noted by \cite{consoliniprl} for
the data seen in Figure 1b is still present in Figure 1d }
\end{figure}

\begin{figure}
  \includegraphics[width=15cm]{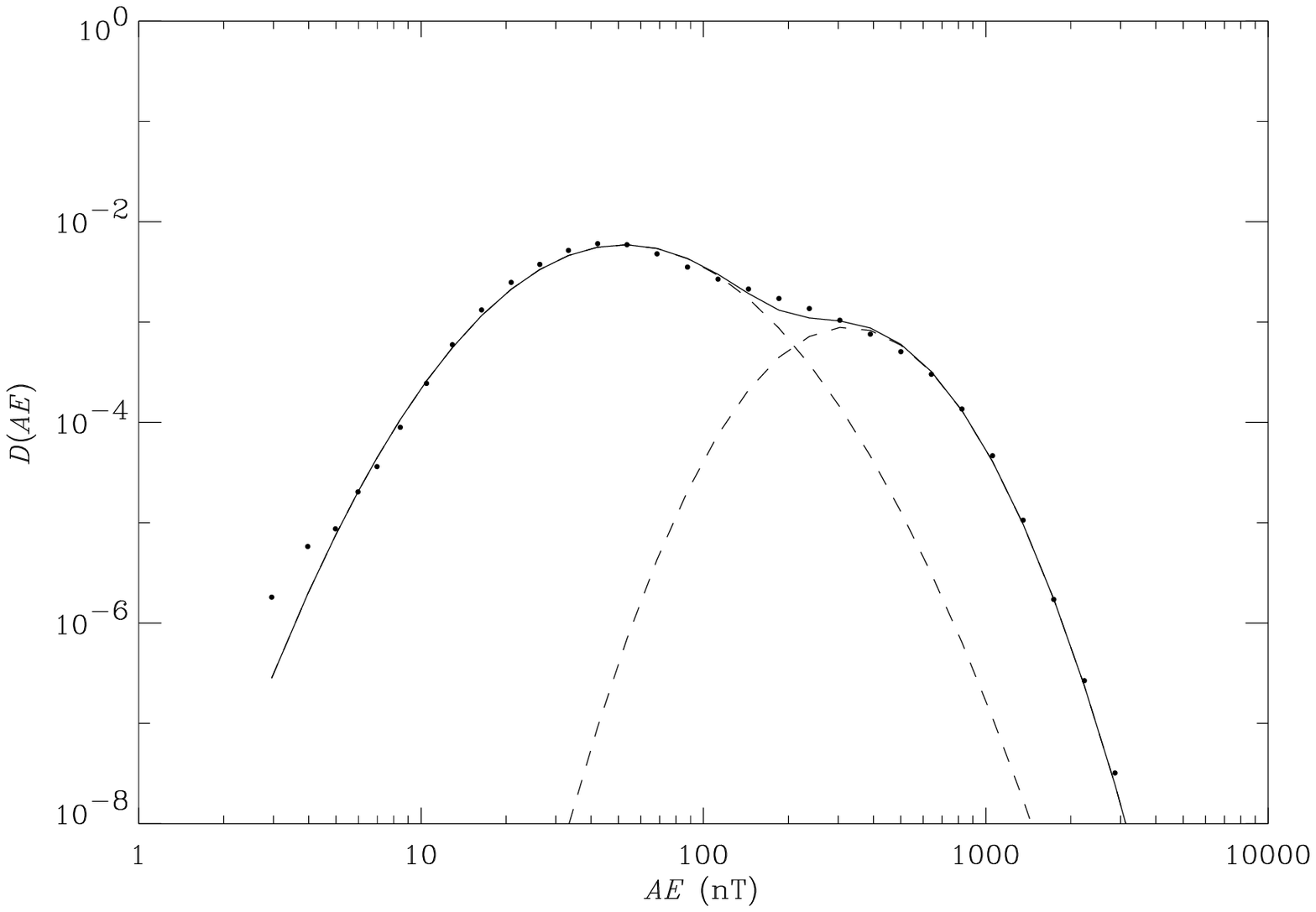}
\caption{ A bi-lognormal fit to 
the PDF of $AE$ for the period January 1978 to June 1978}

    \end{figure}

This conclusion is consistent with the fact that \cite{consolinigrl98} 
found $AE$ to have a two component probability distribution function (PDF),
 well described by  two quasi-lognormal distributions, one of which had an
 exponential cutoff. A similar result was shown by \cite{vassiliadisjgr} for
$|AL|$. In Figure 2  we show that two lognormals,  with the standard prefactor
and without exponential modification, give a very good   fit to the PDF of
$AE$ for the period January 1978 to June 1978.

\subsection{Second order statistics: Fourier spectrum and ACF}

\begin{figure}
  \includegraphics[width=15cm]{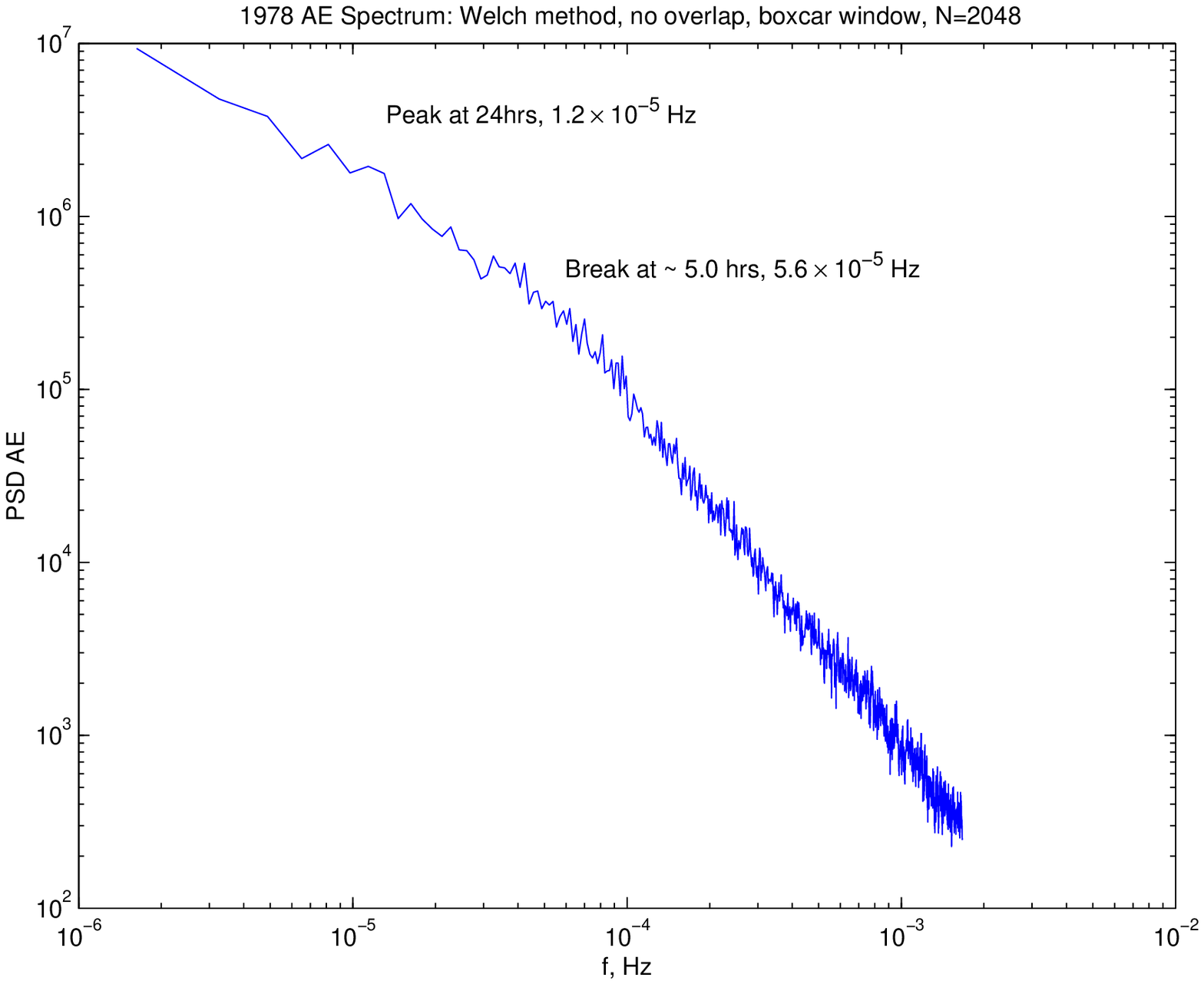}
\caption{PSD for   $AE$ using same approach as that in \cite{tsurutani} but 
for 1 year rather than 3. A similar spectrum is seen over the shorter period,
though the 24 hour peak is not clear in only one year's data.}
    \end{figure}
   
    While the  range of amplitudes present in AE is described by the PDF
     one may study $AE$ in the  frequency domain by Fourier analysis. \cite{tsurutani} used the Welch averaged periodogram 
to obtain a power spectral density for 5-minute $AE$  data 
for the periods 1978-80 and
1967-70. The ``broken power law" spectrum they obtained is typical of $AE$
even for shorter runs of data (see e.g. Figure 3 where Tsurutani et al's
approach is applied to 1 year of $AE$ data). The $1/f$ behaviour is seen essentially all
the time but \cite{consolinieos} has shown that the higher frequency
 $1/f^2$ component is present for high activity levels.
  
 The power spectral density can be described in terms
of two components, one of which (the ``$1/f$" part) 
indicates long range f correlation in 
time.

The autocorrelation function (ACF) of a time series $X(t)$
\begin{equation}
ACF(\tau)= \sum_t X(t) X(t+\tau)
\end{equation}
allows one to express the above behaviour in 
the time rather than the frequency domain. 
\cite{tandt94} found an ACF
with qualitatively two-component behaviour (see Figure 4 which is
 a replotting of their Figure 1 with the data they used). The long-range tail becomes
progressively clearer as one observes longer and longer (20 days or more)
time series, though in  Figure 4 we follow \cite{tandt94} by plotting
only lags up to 500 minutes. Even for short runs of data (5 days), 
however, a fast-dropping exponential ACF is  seen, for which the normalised 
amplitude halves in about 100 minutes. As noted by \cite{mantegna1999} ``fast decaying autocorrelation
 functions and power spectra resembling white noise
 (or ``$1/f^2$" power spectra for the integrated variable) are `fingerprints'
 of short range correlated stochastic processes". Two classic examples of
 such processes are the velocity of a Brownian particle
 (e.g. \cite{mantegna1999}) and the ``random telegraph"
 (a Poisson-switched on-off pulse train (e.g. \cite{jensenbook})).
 
 \begin{figure}
  \includegraphics[width=15cm]{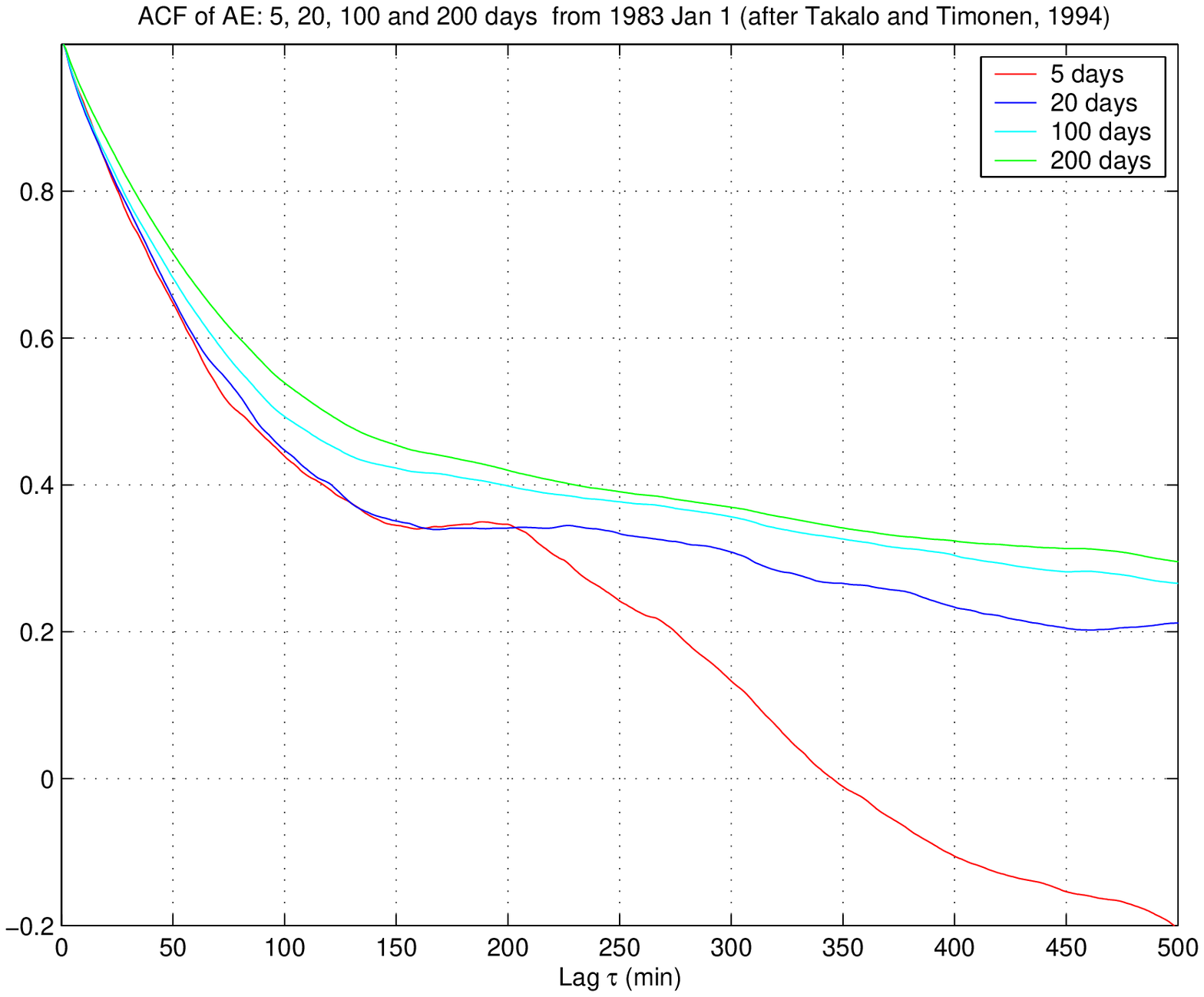}
\caption{ After \cite{tandt94}, comparison of autocorrelation functions
of $AE$ taken over 5, 20, 100 and 200 days in 1983.}

    \end{figure}
 
In order to further examine the time-domain behaviour of the $AE$ series, 
\cite{tandt94} studied the second order structure 
function $S_2(\lambda \Delta t)$. This is defined as
\begin{equation}
S_2(\lambda \Delta t) = < X(t + \lambda \Delta t) - X(t)>^2
\end{equation}
with $<...>$ denoting a time average.
On the assumption of stationarity this equates to $2 (<X(t)>^2 - ACF(\lambda \Delta t) )$.
A self-affine (e.g. \cite{malamud,sornette}) signal has $S_2 (\lambda \Delta t)/S_2 (1) \sim \lambda^{2H}$
(\cite{tandt94}) and so a log-log plot of $(S_2 (\lambda \Delta t)/S_2 (1))^{1/2}$ versus $\lambda$
for such a self-affine signal would give a slope equal to $H$. However, as
we have  discussed elsewhere (\cite{watkins2001b}), an exponential form for the ACF, whatever 
its cause, must necessarily imply  a region of {\bf  $H=0.5$} scaling 
in the structure function. The Taylor series expansion of the exponential
means that the inferred $H$ value must be  $0.5$ (\cite{watkins2001b}), 
and so the nearly  linear scaling region with $H \sim 0.5$ seen by \cite{tandt94}
 might be just another manifestation of the observed
exponential autocorrelation,  rather than in itself necessarily implying that 
the high frequency part of the $AE$ signal is a scale free coloured noise.

\subsection{Burst distributions: Duration, size and waiting time }

The advent of  the idea of self-organised criticality led \cite{takalo1993} and
\cite{consolini1997} to study ``bursts" in the $AE$ index time series. These bursts
were defined using a constant threshold method whereby the set $\{t_1\}$ denotes
all the upcrossings of a given threshold, and the set  $\{t_2\}$ all the downcrossings, 
so burst ``size" $s$ is given by

\begin{equation}
s = \int_{t_1}^{t_2} AE(t) dt
\end{equation}

while ``durations" and ``waiting times" are given by the intervals between
a given $t_1$ and the next $t_2$, and that between a $t_2$ and the next $t_1$
respectively. The first studies of the probability density function (PDF) 
of $s$ were both on a single year's data, and the resulting distributions
of $s$, $\tau$ and $T$ were all power laws with an exponential roll-off. Study of 
longer time series later showed, however (e.g. 
(\cite{consolinieos,freeman2000a}), that the scale-free behaviour was 
interrupted by a bump in both the burst size and duration PDFs. The simplest
interpretation of this is that there are two components in $AE$, presumably the
same two components seen in the PSD and PDF of $AE$ itself, although this 
raises several questions which will be discussed below 
(see also the extended discussions in \cite{freeman2000a,watkins2001b}).

If there are effectively two components to the $AE$ time series then a
natural question is ``where do they come from" ? Because of the long-established
evidence for the driving of the $DP2$ convection currents  (e.g. \cite{kb93})
(believed to be the most continuous contributor to the  $AE$ index) by the solar 
wind,  \cite{freeman2000a} performed an identical constant threshold burst duration 
analysis  on the solar wind $vB_s$ and $\epsilon$ functions, and also 
on the $AU$ and $AL$ indices. It was found that burst lifetimes in all four quantities showed 
power laws with very similar slopes. However, the aforementioned
bump  in the burst duration signal was seen  only in the magnetospheric signals. Because
the bump was larger in $|AL|$  (mainly drawn from post-midnight stations) than $AU$ (mainly
dusk), \cite{freeman2000a} inferred that the bump was the signature of the
substorm  $DP1$ current. The  inference drawn from this study was that the scale-free component in $AE$ was most probably
a consequence of scale-free behaviour in the solar wind, and that the `bump" was an objective
identification of the substorm with a characteristic time scale.

A number of important criticisms may be raised of the result, however. Obviously it
was not ideal to use non-contemporary time series (1978-88 for the $AU$ and $AL$ indices and
1995-98 for the solar wind data). 
 Unfortunately no 2 simultaneous datasets of 
many years' length were available for $AU/AL$ and the solar wind data.
This problem was addressed in part by comparison between the 
burst duration PDFs taken from a 4-year period of the $AU$ and $AL$ series and the existing solar
wind burst PDFs from an equivalent phase of the solar cycle. No significant difference was
seen between the burst PDFs taken from this section and those from a full 10 year series of $AE$.
The main justification for the approach used, however, is the remarkable 
stability of the {\bf bi-lognormal aspect of }underlying PDF of $AE$. 
While the curves vary in 
parameters, they can consistently be fitted by bi-lognormal
fits. 
In consequence, if as we suspect, the power laws found are partially
a consequence of the ``1/x" power law scaling region present
in a lognormal distribution (\cite{sornette}) regardless of its parameters,  
they will be stable from year to year.   Each curve in figure 5 is the PDF of of $AE$ taken over one of the years
from 1978 to 1988. A similar stability was found by \cite{burlaga2000}
in PDFs of the solar wind velocity, density and temperature during the 
period from which the WIND measurements were taken.

\begin{figure}
  \includegraphics[width=15cm]{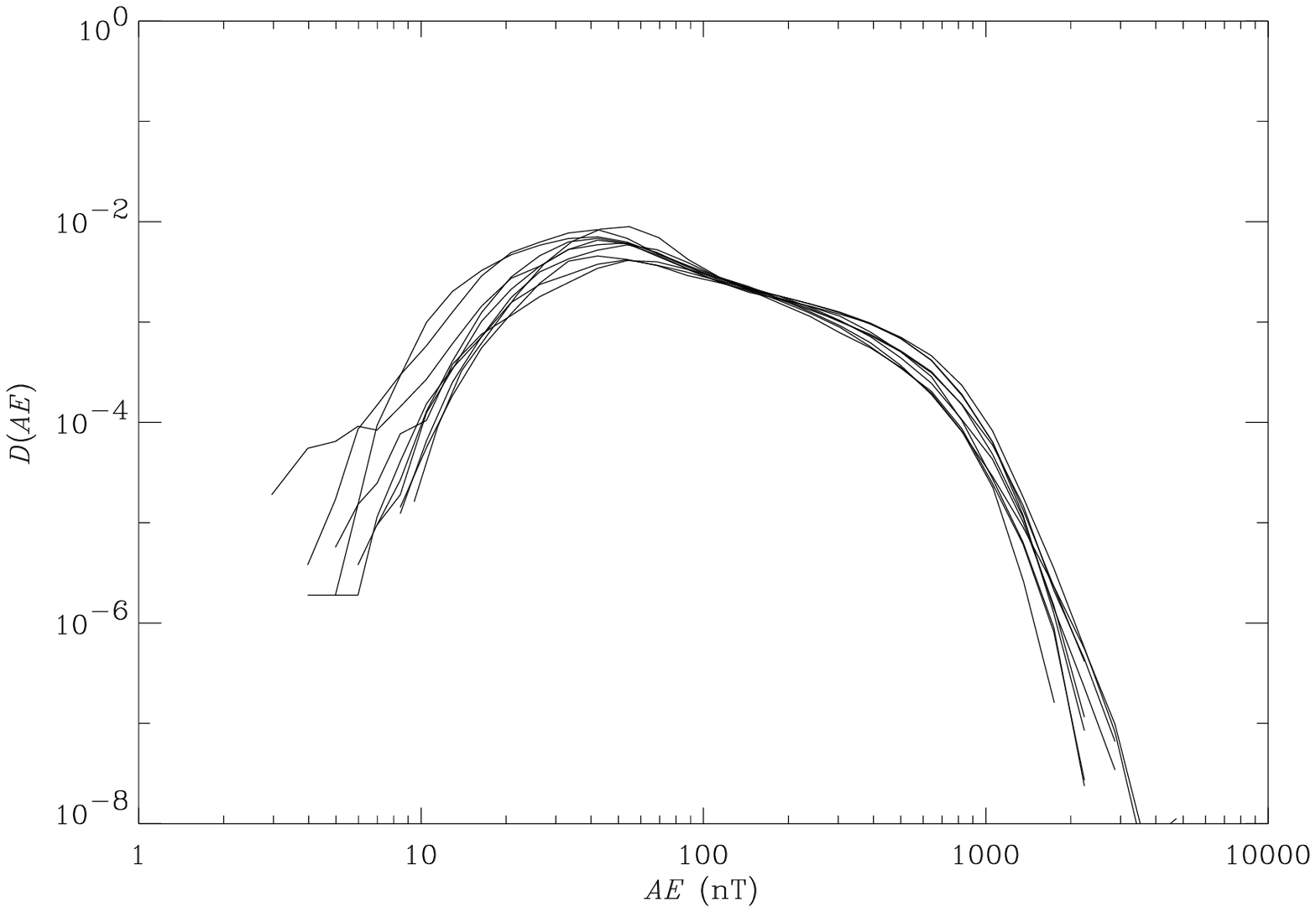}
\caption{ Each curve   is the PDF of $AE$ taken over one of the years
from 1978 to 1988.}

    \end{figure}

In addition, to eliminate the effect of the angle $\theta$ which represents
the solar wind-magnetosphere coupling process in Akosofu et al's $\epsilon$
function, \cite{freeman2000b} subsequently examined the PDFs for burst durations,
waiting times and sizes in the radial solar wind Poynting flux. Both durations
and waiting times could be fitted by identical exponentially-truncated
power laws, while the apparent power law region in burst size is
also reminiscent of  long-ranged lognormal distributions (\cite{malamud,sornette}).

More recently, \cite{uritskygrl} have examined a    
 contemporary, pair of datasets for solar wind $\epsilon$ and $AE$
(although the series are shorter (6 months) and 5 minute resolution). 
They find different scaling behaviours for time-averaged  activity and event survival probability
in the two signals,  and an apparent break in the scaling of the dependence
of $AE$ burst size on duration. They interpreted these results  as evidence
of a distinct internal magnetospheric dynamical component in the $AE$ signal.

\subsection{Fractional lognormal motion and ``persistent lognormality"}
 It is straightforward to illustrate
how  some of the above conjectures can be checked  by using a
numerical realisation of  fractional lognormal noise. The details will be presented
elsewhere but we here sketch how such a process may be used to approximate the
``1/f" component of the AE series. We postpone to a subsequent work implementation of
a suitable model of the second, unloading, component that has been postulated above,
but note that a first order approximation to the ``1/f$^2$" term was presented in \cite{watkins2001b}.
We made a time series of fractional lognormal motion with zero mean and unit 
standard deviation, and spectral exponent $-1.5$, using the algorithm of \cite{malamud}.
Taking the spacing of the data to be 1 minute, we plot 
4000 points from the resulting time series
in Figure 6.  From a longer series we have confirmed  that power
law PDFs for  size, duration and waiting times of bursts
as defined in e.g. \cite{consolini1997} do indeed result from such a model,
while we also see size-duration relations, with  scale breaks, qualitatively
similar to those seen by \cite{uritskygrl}.

The power spectrum and amplitude distribution of $Y$ are, as
expected, $f^{-1.5}$ and lognormal, while perhaps
more surprisingly, the differenced time series $X$ has
leptokurtic tails and shows a range over which a scaling collapse 
of the type described by \cite{chapman2002} can be demonstrated. However 
agreement with with observed autocorrelation functions and Hurst exponent measures
is relatively poor, confirming  that we need to include a representation of the unloading 
component before a useful comparison can be made.

We note that fractional lognormal motion has several competitors as a model  
of real anaomalous diffusion, most notably truncated Levy
flights, and continuous time Levy flights, the latter being generalised
random walks in which Levy stable increments are combined with a random
distribution of times between steps \citep{paul}. We plan to investigate these alternatiives
 to find an optimum description of the driven component of AE.
 
\begin{figure}
  \includegraphics[width=15cm]{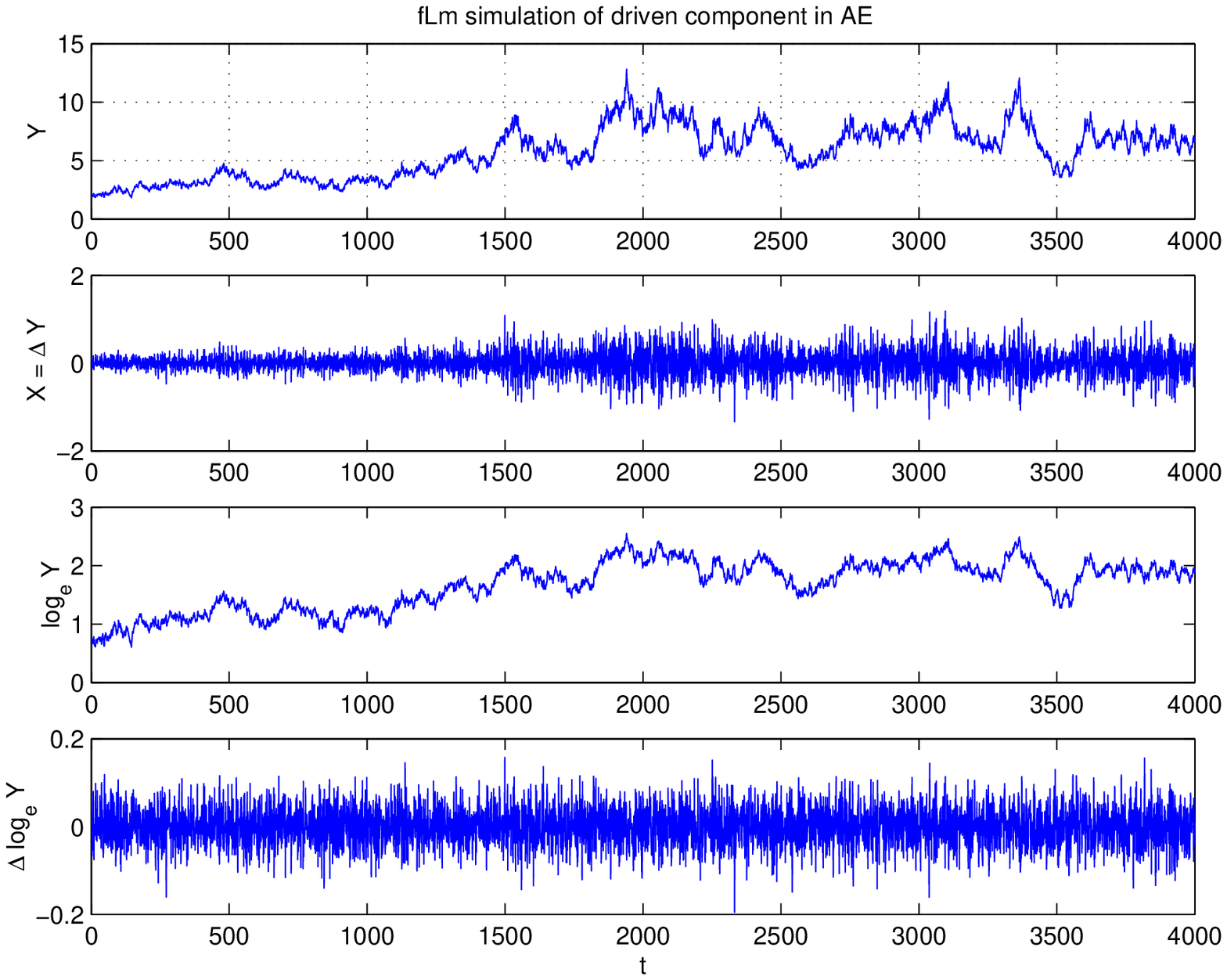}
\caption{The top panel shows a realisation of 4000 points of a synthetic AE ``driven component" Y,
with lognormal amplitude distribution and a power spectral density
$\sim f^{-1.5}$. The second panel shows the first difference (compare figure 1b),
the third panel shows the natural log of Y (compare figure 1c), while
the fourth panel shows the differenced log Y time series, in which the
intermittency apparent in the second panel has disappeared. }    
\end{figure}

\subsection{Summary of the ``stylised facts" of $AE$}

We may summarise the above non-exhaustive collection of ``stylised facts" about $AE$,
with some tentative interpretations,  as follows. 

\begin{enumerate}

\item  The $AE$ time series has  more than one component (see also e.g. Table 5.2 in \cite{kb93}).

\item If $AE$ has more than one component it may not necessarily mean that more than
one process is at work. For example, \cite{chapman1998} showed that a sandpile 
model could produce both a scale-free and non-scale-free component, where the 
non -scale-free component is identified with the systemwide events in the model).

\item At least one of the components in $AE$ is long-range correlated (and fractal)
in time.

\item This observed long range correlation may originate in that  
{\bf  present } in the solar wind Poynting flux, a hypothesis supported by the identical
power laws seen both in waiting time and burst duration by
\cite{freeman2000b}, as expected for the isosets  of a fractal  random walk
 (e.g. \cite{watkins2001a}).

\item One component apparently has a characteristic time scale, which
\cite{freeman2000a} identify with the substorm.

\item The PDF of the $AE$ signal itself is apparently not scale free, but rather
is well described by a bi-lognormal. However it is already known that such 
distributions are also very good approximations to underlying
multifractal cascades (see e.g. the discussion in \cite{burlaga2001}), and so, because of
evidence for multifractality in $AE$ (e.g. \cite{consoliniprl}), 
we admit that a fractional lognormal noise
component may be a less well motivated choice of model for the 
long range correlated part of $AE$ than it might at first seem.

\item Nonetheless it is possible that the power law PDF for burst durations and
waiting times results simply because one part of the $AE$ signal is
well approximated by fractional lognormal noise. In consequence the less clear cut scaling
in burst size would then result from the convolution 
(implicit in the method used to define bursts) between the non scale-free
amplitude PDF and  the genuinely scale-free persistent property of
 the $AE$ time series.  

\end{enumerate}

\section{Outstanding questions}

The long range time correlation and long tailed PDFs seen in $AE$ can be physically
motivated by sporadic, localised energy release events in the magnetotail,
and can be modelled by  sandpile algorithms (e.g. \cite{chapman1998}). The observed
2-component property discussed above can also be accounted for; at least if
we make the hypothesis of a difference between systemwide and internal energy release 
events. However, a number of issues arise from the summary presented above,
and it is still the view of the present author that more definitive answers
to the following questions are needed, to establish more clearly if SOC is 
a necessary explanation for the behaviour of $AE$ rather than a possible one.

\subsection{Does time correlation imply SOC?}

The answer to this would seem to be no, in that one may have long-range
correlation in time, signalled by ``$1/f$" spectra, in many other classes 
of process, some of which are not spatially long range correlated at all.
 Conversely, some (but not all) SOC models
(most notably the orginal BTW model (\cite{btw87})), do not exhibit long 
range time correlation in their outputs. In 
consequence, it is neither advisable  to test for SOC relying only
on ``$1/f$" spectra nor to use the power law waiting times seen in shell
models as a test against SOC  (see also \cite{watkins2001a} and references therein,
where this issue is discussed in reference to solar flares, and the
discussion in the context of fusion plasmas of \cite{krommes}).

\subsection{Is the observed amplitude (rather than burst) PDF in $AE$
consistent with criticality? }

The problem with interpreting power laws in burst size as evidence of scale free
behaviour is that one may also have a time-correlated lognormal process
(see e.g. the simulations presented in Figure 19 of \cite{malamud}, and
the illustrative time series shown in figure 6). 
Integrating a times series of the amplitude of such a process in the region
where it mimics a power law (e.g. Figures 4.2 and 4.3 
in \cite{sornette})  gives rise to a power-law like burst distribution, provided that the signal is persistent in time.
In fact even burst distributions constructed from a Gaussian persistent process
are remarkably power-law like (see figure 2 of \cite{watkins2001a}).
We thus currently need to consider ``persistent lognormality" as well as
SOC as an alternative explanation for the stylised facts of $AE$ bursts.

\subsection{Can power spectra be reconciled with burst distributions?}

It has been remarked [Channon Price, personal communication, 1999]
that the monofractal nature of the burst  duration distributions seen would seem hard to reconcile with the
 apparent biaffine nature of the power spectrum of $AE$. This is because a
 mono-fractal (or bifractal) should give the same H value (or values)
 to both 1st order and 2nd order measures. The resolution of
 this apparent paradox is that a mixed signal, which has a fractal
 part and a non-fractal part does not have this
 limitation. A first order measure like burst duration basically
 measures the length of a fractal curve (because the slope of the 
 isoset distribution is governed by the fractal dimension of the curve which is
 crossing the threshold). If a signal has a fractal component which
 accounts for most of the curve, it may take a long time series for the 
 non-fractal part to become apparent. In the author's view this is the
 reason why the bump in $AE$ burst sizes needed a long time
 series (as used by \cite{consolinieos}) to become apparent, rather than the 
 1-year series studied by  \cite{consolini1997}.
 
 A second order measure such as $S_2$, conversely, measures the distribution
 of variance in a signal, because of the formal equivalence between the
 information in $S_2$ and that in the ACF or power spectrum. A sharp change
 in level in the time series caused by a non-fractal substorm component
 can be revealed by the $1/f^2$ power spectrum it gives rise to at high frequencies.
 The presence of such a non-fractal component explains why, in contrast with first order methods,
 the power spectrum reveals a spectral break in even as few as 4 days of $AE$ data,
 and it is clearly apparent in Figure 3 taken from 1 year of $AE$ data.

\subsection{How much of the time correlation in $AE$ comes from the solar wind?}

It seems to be increasingly accepted (e.g. \cite{takalo1999,price2001,uritskygrl})
 that some part of the scale-free behaviour of
the $AE$ signal comes from the scale-free solar wind driver. However 
two caveats need to be noted. One is that, as with $AE$, the solar wind 
probably also has a multifractal character, and so comparison of fractal dimensions
as in \cite{freeman2000a} between the solar wind and AU/AL is really just 
comparing measures at one order. However, \cite{voros1998}
 showed that the
higher order structure functions of low pass filtered, solar wind magnetic field fluctuations 
and those for an unfiltered ground-based magnetometer signal were {\bf also}
in close agreement. More importantly, the Hurst exponent for a signal is its
roughness averaged over many length scales. Even if the $AE$ output is 
 nonlinearly driven by the solar wind, similar roughness values in the output
 and the driver  would not require individual bursts to be the same. 
 In consequence
 the absence of one to one mappings between integrated input power
 and $AE$ bursts, as found  by \cite{uritskygrl},
 may not be so surprising. We will return to
 this point in future work.

\section{Further motivation for criticality or SOC behaviour}

For all the above reasons, better indicators of SOC behaviour than
those derived from $AE$ are needed in the magnetospheric
case.  \cite{uritskyjgr} have recently shown remarkably clear power laws in several
burst measures drawn from time-evolving ``blobs" seen by the UVI auroral imager on
the WIND satellite. While more studies are necessary, a particularly interesting
result is the $E^{-1.5}$ power law that they observe in the time integrated energy
of blobs.  Following \cite{uritskyjgr} in assuming   that the blobs they 
observed in UVI data correspond to magnetic reconnection events in the magnetotail, it is tempting to explain the
observed energy spectrum using the results of  \cite{craig}. He was
 considering simple models of solar flares and showed that 
an $E^{1.5}$ power law is exactly the spectrum to be expected from distributed  reconnection
events with 2D current sheet geometry, 
 using the known properties of analytic solutions for
reconnection and a relatively conservative set of additional assumptions.
The optical observations of \cite{uritskyjgr} thus suggest SOC or least
more obviously SOC-like behaviour more directly than $AE$. They illustrate
clearly how by studying different physical quantities or measures will
help better answer the question of whether SOC is present in the magnetosphere.

\section{Conclusions}

SOC is not yet needed to explain the properties of $AE$'s  
time series: at least where the PDF, power spectrum, structure function
and ``burst statistics" are concerned. Nonetheless, because SOC  is an economical
and physically motivated approach to capturing these properties,
further studies are needed to clarify some outstanding issues, in particular
the relationship between solar wind scaling and that seen in $AE$.
A recent pointer to how the SOC investigations may be advanced is
the demonstration of power law scaling in spatiotemporal events in
a signal (UVI images) which may be less ambiguous than the necessarily ``mixed"
signal seen by $AE$.

\begin{acknowledgements}
This paper is based on an invited talk given at the IAGA  meeting in
Hanoi, Vietnam, August 2000. NWW thanks the conveners of this session for the opportunity
to give this talk, and their dedicated efforts at the meeting.
He would also like to thank Mervyn Freeman, and the participants in the
2002 Venice Workshop on Complexity in the Earth's Magnetospheric
Dynamics  for innumerable 
stimulating discussions; David Riley for his work in preparing Figures 2 and 5;
and Vadim Uritsky and Alex Klimas for a preprint of \cite{uritskyjgr}.
\end{acknowledgements}

\bibliographystyle{egs}

\end{document}